\documentclass[trackchanges]{aastex701}
\usepackage{siunitx}
\usepackage{booktabs}
\usepackage{amsmath}

\begin{document}

\title{The Formation Rate and Luminosity Function of Fast X-ray transients from Einstein probe}
\author{Yizhou Guo}
\affiliation{Purple Mountain Observatory, Chinese Academy of Sciences, Nanjing 210023, People's Republic of China}
\affiliation{School of Astronomy and Space Science, University of Science and Technology of China, Hefei 230026, People's Republic of China}
\email{} 

\author[orcid=0000-0001-8500-0541]{Houdun Zeng}
\affiliation{Purple Mountain Observatory, Chinese Academy of Sciences, Nanjing 210023, People's Republic of China}
\affiliation{School of Astronomy and Space Science, University of Science and Technology of China, Hefei 230026, People's Republic of China}
\affiliation{Key Laboratory of Astroparticle Physics of Yunnan Province, Yunnan University, Kunming 650091, People's Republic of China}
\email[show]{zhd@pmo.ac.cn}

\author[orcid=0000-0003-0162-2488]{Junjie Wei}
\affiliation{Purple Mountain Observatory, Chinese Academy of Sciences, Nanjing 210023, People's Republic of China}
\affiliation{School of Astronomy and Space Science, University of Science and Technology of China, Hefei 230026, People's Republic of China}
\email[show]{jjwei@pmo.ac.cn}

\author[orcid=0000-0003-2915-7434]{Hao Zhou}
\affiliation{Purple Mountain Observatory, Chinese Academy of Sciences, Nanjing 210023, People's Republic of China}
\affiliation{School of Astronomy and Space Science, University of Science and Technology of China, Hefei 230026, People's Republic of China}
\email{} 

\author[orcid=0000-0003-4977-9724]{Zhiping Jin}
\affiliation{Purple Mountain Observatory, Chinese Academy of Sciences, Nanjing 210023, People's Republic of China}
\affiliation{School of Astronomy and Space Science, University of Science and Technology of China, Hefei 230026, People's Republic of China}
\email{} 

\author[orcid=0000-0002-6299-1263]{Xuefeng Wu}
\affiliation{Purple Mountain Observatory, Chinese Academy of Sciences, Nanjing 210023, People's Republic of China}
\affiliation{School of Astronomy and Space Science, University of Science and Technology of China, Hefei 230026, People's Republic of China}
\email[]{}

\author[orcid=0000-0002-9758-5476]{Daming Wei}
\affiliation{Purple Mountain Observatory, Chinese Academy of Sciences, Nanjing 210023, People's Republic of China}
\affiliation{School of Astronomy and Space Science, University of Science and Technology of China, Hefei 230026, People's Republic of China}
\email[show]{dmwei@pmo.ac.cn} 

\begin{abstract}

Following its launch on 2024 January 9, the Einstein Probe (EP) telescope has detected hundreds of fast X-ray transients (FXTs), yet their physical origins remain elusive. Understanding their luminosity function and formation rate is crucial for elucidating their nature. Recently, the EP team has provided the latest catalog of EP-detected FXTs \citep{Wu2025_inprep}. Based on this catalog, we present a model-independent nonparametric approach to derive the luminosity function and formation rate of FXTs.
Our analysis reveals significant cosmological luminosity evolution, characterized by a scaling relationship of $(1+z)^{3.58}$. After accounting for this evolution, we establish that the local luminosity function is best represented by a broken power law, with a break luminosity of $(4.17 \pm 0.34) \times 10^{46}$ erg/s. The formation rate exhibits a broken power law as $\rho(z) \propto (1+z)^{-4.25}$ at $z \lessapprox 0.9$ and $\rho(z) \propto (1+z)^{-0.26}$ at $z \gtrapprox 0.9$, yielding a local rate of approximately $153.8_{-95.1}^{+249.4}$ Gpc$^{-3}$ yr$^{-1}$. This rate is higher than that of long gamma-ray bursts (LGRBs). Our findings indicate that a component of FXTs is associated with LGRBs.
\end{abstract}

\keywords{\uat{X-ray astronomy}{1810} --- \uat{X-ray transient sources}{1852} --- \uat{Gamma-ray bursts}{629}}

\section{INTRODUCTION}
Fast X-ray transients (FXTs) are brief, intense bursts of soft X-ray emission in the $0.5-4$ keV range, typically lasting from minutes to hours. Although these events were first detected several decades ago, their origins remain unclear. Early research suggested that FXTs might be linked to gamma-ray bursts (GRBs), particularly based on the detection of prompt X-ray emissions by missions like Ginga, GRANAT/WATCH, and HETE-2 \citep{1998ApJ...500..873S,1998A&AS..129....1S,2004ApJ...617.1251V}. Additionally, the BeppoSAX mission helped identify X-ray afterglows associated with GRBs \citep{1997Natur.387..783C,1998A&A...331L..41P}. 
However, these narrow-field X-ray telescopes are insufficient for rapidly increasing the sample size of FXTs available for research.

The launch of the Einstein Probe (EP) in 2024 January marked a significant step forward in the study of FXTs \citep{Yuan2022,2025SCPMA..6839501Y}. EP's wide-field X-ray telescope (WXT) has significantly increased the detection rate of these transients, revealing a variety of emission properties. A key discovery was the detection of EP240315a, the first FXT with confirmed optical and radio counterparts, which is strongly associated with a long GRB (LGRB; \citealt{liu2024softxraypromptemission,Gillanders_2024}). This high-redshift event ($z = 4.859$) and its optical counterpart, a Type Ic supernova, suggest a possible link between FXTs and GRBs \citep{van_Dalen_2025,Srivastav_2025}. However, not all FXTs observed by EP show gamma-ray emissions, challenging the idea that all FXTs are GRB related. Examples such as EP240408A, EP240414a, and EP241021a, which lack detectable gamma rays, further complicate the picture \citep{2025SCPMA..6819511Z,O’Connor_2025,van_Dalen_2025,Srivastav_2025}. Some of these events exhibit X-ray properties similar to GRB afterglows, raising questions about whether FXTs are related to GRBs or represent a distinct class of astrophysical phenomena.

Despite these uncertainties, the growing body of data—especially from EP—provides a unique opportunity to better understand FXTs. 
\citet{2025arXiv250907141O} used 26 public EP/WXT transients with secure spectroscopic redshifts to analysis their redshift distribution and found it is consistent with that of long-duration GRBs.
Recently, the EP team has released a comprehensive catalog of FXTs, containing 107 events as of 2025 August 31 \citep{Wu2025_inprep}.
This dataset offers a valuable chance to conduct a comprehensive statistical analysis, free from the biases that typically arise from different observational instruments. The objective of this study is to derive their luminosity function and formation rate in a nonparametric manner base on the sample of FXTs observed by EP \citep{Wu2025_inprep}. By investigating the statistical characteristics of these transients, we aim to provide new insights into their underlying mechanisms and explore their potential relationship with GRBs and other high-energy transients.

The paper is organized as follows: Section~\ref{sec:sample} describes the data sample, and in Section~\ref{sec:function} we present a brief description of the nonparametric statistical method and apply it to derive the luminosity function and formation rate. The conclusion and discussion are presented in Section~\ref{sec:conclusion}. Throughout this paper a standard $\Lambda$CDM cosmology \citep{2020A&A...641A...6P} with $H_0=67.4$ km $s^{-1}$ Mpc$^{-1}$ and $\Omega_m=0.315$ is adopted.

\section{data SAMPLE} \label{sec:sample}

\citet{Wu2025_inprep} present a uniformly processed catalog of EP-detected FXTs up to 2025 August 31, comprising 107 events. It provides consistent EP X-ray light curves and spectral products, specifically the photon index and time-averaged X-ray flux, and also it compiles multiwavelength metadata such as redshifts. From this catalog, we adopt a subsample of 31 FXTs with secure redshifts. 
The properties of this subsample are listed in Table~\ref{tab:sample}.

\begin{figure*}[ht!]
\plotone{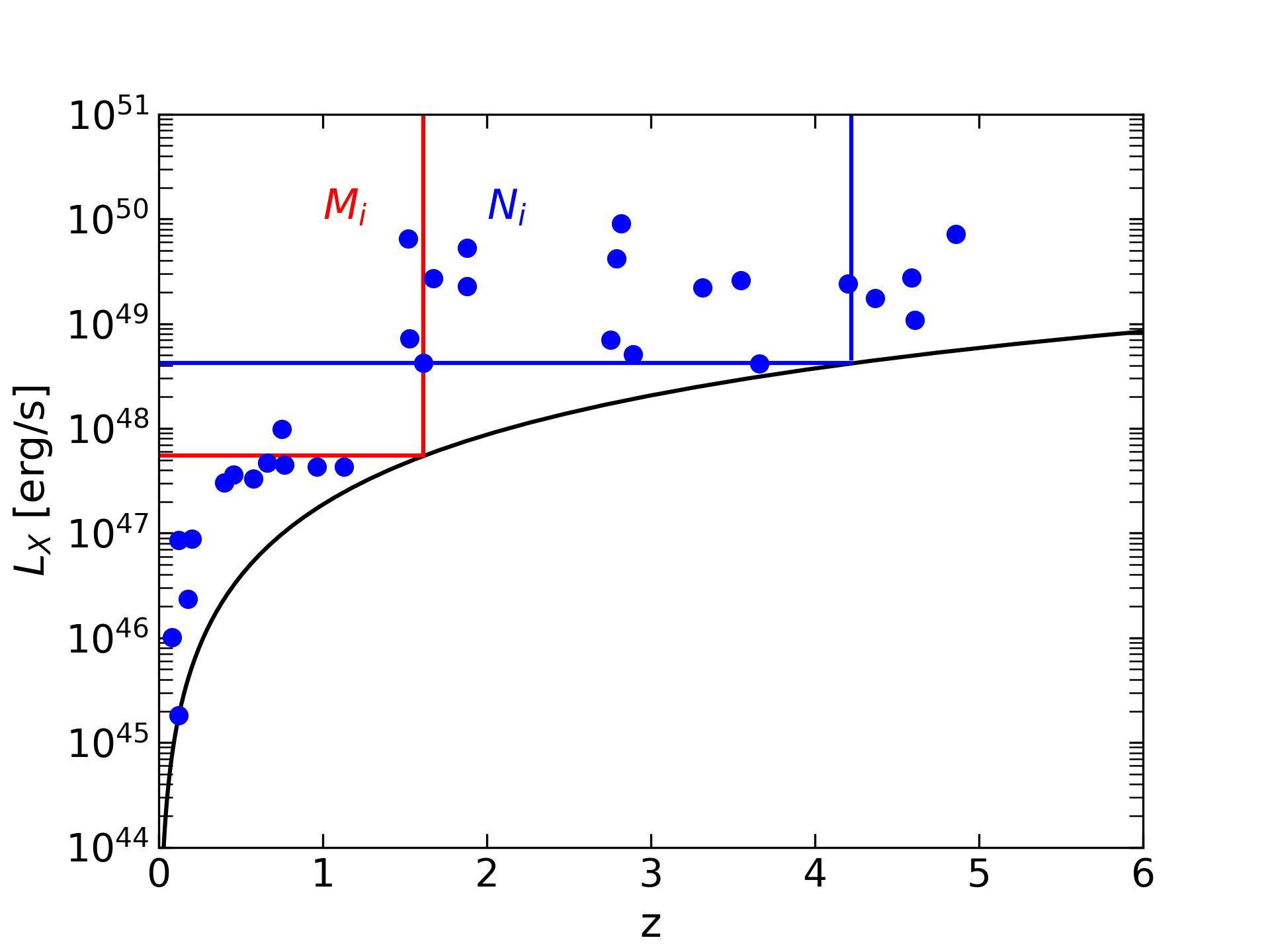}
\caption{The X-ray luminosity vs. redshift of 31 FXTs. The solid (black) curve shows the truncation boundaries assuming limiting X-ray flux of $F_{\rm limit} = 4.5 \times 10^{-11}$ erg cm$^{-2}$ s$^{-1}$ with a fixed photon index $\Gamma_{\rm WXT}= 1.59$.  
The associated set $M_i$ is shown in the red rectangle, and $N_i$ is shown in the blue rectangle.
\label{lum_z}}
\end{figure*}

The X-ray luminosity of FXTs can be calculated in the rest frame $0.5-4$ keV by
\begin{equation}
L_{X}=\frac{4\pi d^2_L(z)F_{\rm WXT,avg}}{(1+z)^{2-\Gamma_{\rm WXT}}} \,,
\label{eq:Lx}
\end{equation}
where $d_L(z)$
is the luminosity distance at redshift $z$, $F_{\rm WXT,avg}$ is the time-averaged X-ray flux, $\Gamma_{\rm WXT}$ is the photon index, and $(1+z)^{2-\Gamma_{\rm WXT}}$ is the usual $K-$correction term. 
In Figure \ref{lum_z}, the blue dots show the X-ray luminosity versus redshift for the 31 FXTs.
The black curve shows the limiting luminosity derived from Equation \ref{eq:Lx}, where $F_{\rm WXT,avg}$ is replaced by $F_{\rm limit} = 4.5 \times  10^{-11}$ erg cm$^{-2}$ s$^{-1}$,  adopting an average spectral index of $\Gamma_{\rm WXT}= 1.59$ obtained from our sample.
Here, $F_{\rm limit}$ is the lowest measured flux value in our sample, which is also the lowest flux limit from the 107 events detected by EP. It therefore represents the detection sensitivity for the faintest source.

\begin{deluxetable*}{lllcccc}
\tablecaption{The sample of EP sources used in this work selected from the latest catalog of EP-detected FXTs provided by \citet{Wu2025_inprep}.\label{tab:sample}}
\tablewidth{700pt}
\tablehead{
\colhead{Name} & \colhead{R.A.} & \colhead{Decl.} & \colhead{GRB} & \colhead{Redshift} & \colhead{$\Gamma_{\mathrm{WXT}}$} & \colhead{$F_{\mathrm{WXT,avg}}$}  \\ 
\colhead{} & \colhead{(deg)} & \colhead{(deg)} & \colhead{} & \colhead{} & 
\colhead{}  & \colhead{erg cm$^{-2}$ s$^{-1}$} 
} 
\startdata
EP240315a	&	141.639 	&	-9.541 	&	GRB 240315C	&	4.859	&	$1.62_{-0.22 }^{+0.22}$ &	$5.56_{-0.43}^{+0.48} \times 10^{-10}$\\
EP240414a	&	191.510 	&	-9.716 	&		&	0.4007	&	$2.33_{-0.53}^{+0.59 	}$&	$4.50_{-0.87}^{+1.43} \times 10^{-10}$\\
EP240506a	&	213.973 	&	-16.707 	&		&	0.12	&	$0.77_{-0.41}^{+0.42}$&	$2.44_{-0.53}^{+0.67} \times 10^{-9}$\\
EP240801a	&	345.127 	&	32.598 	&	XRF 240801B	&	1.673	&	$1.41_{-0.08}^{+0.09}$&	$2.45_{-0.13}^{+0.14} \times 10^{-9}$\\
EP240804a	&	337.649 	&	-39.099 	&	GRB 240804B	&	3.662	&	$1.34_{-0.20}^{+0.21}$&	$8.74_{-1.20}^{+1.31} \times 10^{-11}$\\
EP240806a	&	11.491 	&	5.091 	&		&	2.818	&	$2.15_{-0.46}^{+0.50}$&	$1.06_{-0.17}^{+0.24} \times 10^{-9}$\\
EP241021a	&	28.852 	&	5.957 	&		&	0.75	&	$1.48_{-0.72}^{+0.73}$&	$4.80_{-1.65}^{+3.12} \times 10^{-10} $\\
EP241025a	&	333.763 	&	83.566 	&	GRB 241025A	&	4.2	&	$1.29_{-0.08}^{+0.08}$&	$4.30_{-0.20}^{+0.21} \times 10^{-10}$\\
EP241026a	&	293.399 	&	57.998 	&	GRB 241026A	&	2.79	&	$1.47_{-0.37}^{+0.39}$&	$1.25_{-0.16}^{+	0.19} \times 10^{-9}$\\
EP241107a	&	35.016 	&	3.323 	&		&	0.456	&	$1.71_{-0.09}^{+0.09}$&	$4.87_{-0.24}^{+0.25}  \times 10^{-10}$\\
EP241113a	&	131.989 	&	52.377 	&		&	1.53	&	$1.46_{-0.20}^{+0.19 }$&	$7.61_{-0.91}^{+1.07} \times 10^{-10}$\\
EP241217a	&	46.950 	&	30.920 	&		&	4.59	&	$1.60_{-0.27}^{+0.33}$&$	2.47_{-0.40}^{+0.43} \times 10^{-10}$\\
EP241217b	&	84.159 	&	-25.299 	&	GRB 241217A	&	1.879	&	$1.51_{-0.04}^{+0.04}$&	$1.47_{-0.03}^{+0.03} \times 10^{-9}$\\
EP250108a	&	55.623 	&	-22.509 	&		&	0.176	&	$3.65_{-1.19}^{+1.62}$&$	1.93_{-0.97}^{+4.98} \times 10^{-10}$\\
EP250125a	&	175.369 	&	-21.707 	&		&	2.89	&	$1.20_{-0.22}^{+0.23}$&$	2.03_{-0.30}^{+0.34} \times 10^{-10}$\\
EP250205a	&	113.522 	&	32.363 	&	GRB 250205A	&	3.55	&	$2.48_{-0.24}^{+0.25 }$&	$1.03_{-0.12}^{+0.13} \times 10^{-10}$\\
EP250207b	&	167.495 	&	-7.906 	&		&	0.082	&	$0.56_{-0.48}^{+0.49 }$&	$6.36_{-1.74}^{+2.36} \times 10^{-10}$\\
EP250215a	&	156.335 	&	-27.717 	&	GRB 250215A	&	4.61	&	$1.32_{-0.42}^{+0.45}$&	$1.56_{-0.22}^{+0.27} \times 10^{-10}$\\
EP250223a	&	98.278 	&	-22.442 	&		&	2.756	&	$1.70_{-0.20}^{+0.20 }$&	$1.59_{-0.18}^{+0.20} \times 10^{-10}$\\
EP250226a	&	224.254 	&	20.982 	&	GRB 250226A	&	3.315	&	$1.08_{-0.14}^{+0.14 }$&	$8.21_{-0.78}^{+0.85} \times 10^{-10}$\\
EP250302a	&	169.499 	&	33.577 	&		&	1.131	&	$1.39_{-0.18}^{+0.19 }$&	$9.10_{-1.06}^{+1.16} \times 10^{-11}$\\
EP250304a	&	208.388 	&	-42.817 	&		&	0.2	&	$2.28_{-0.06}^{+0.06 }$&	$6.79_{-0.23}^{+0.23} \times 10^{-10}$\\
EP250321a	&	179.256 	&	17.370 	&	GRB 250321D	&	4.368	&	$0.89_{-0.20}^{+0.21 }$&	$5.69_{-0.40}^{+0.43} \times 10^{-10}$\\
EP250404a	&	125.056 	&	35.518 	&	GRB 250404A	&	1.88	&	$1.62_{-0.04}^{+0.04 }$&	$3.02_{-0.06}^{+0.06} \times 10^{-9}$\\
EP250416a	&	256.409 	&	25.769 	&	GRB 250416C	&	0.963	&	$1.39_{-0.24}^{+0.25 }$&	$1.29_{-0.16}^{+0.18} \times 10^{-10}$\\
EP250427a	&	277.281 	&	7.570 	&	GRB 250427A	&	1.52	&	$1.76_{-0.19}^{+0.19 }$&	$5.20_{-0.39}^{+0.46} \times 10^{-9}$\\
EP250430a	&	233.399 	&	-18.130 &	GRB 250430A	&	0.767	&	$1.74_{-0.43}^{+0.47}$&	$1.81_{-0.25}^{+0.34} \times 10^{-10}$\\
EP250704a	&	300.878 	&	12.007 	&	GRB 250704B	&	0.661	&	$1.83_{-0.18}^{+0.19 }$&	$2.56_{-0.23}^{+0.25} \times 10^{-10}$\\
EP250821a	&	289.878 	&	-43.394 	&		&	0.577	&	$1.60_{-0.08}^{+0.08}$&	$2.75_{-0.13}^{+0.14} \times 10^{-10}$\\
EP250827a	&	3.484 	&	-56.468 	&		&	1.613	&	$0.71_{-0.22}^{+0.22 }$&	$8.13_{-1.18}^{+1.36} \times 10^{-10}$\\
EP250827b	&	36.581 	&	37.499 	&		&	0.12	&	$2.04_{-0.43}^{+0.48}$&	$4.51_{-1.01}^{+1.28} \times 10^{-11}$\\
\enddata
\tablecomments{The X-ray photon index $\Gamma_{\rm WXT}$ and time-averaged X-ray flux are reported in the $0.5- 4$ keV band. Moreover, 15 events in our sample were also detected as GRBs. }
\end{deluxetable*}

\section{LUMINOSITY FUNCTION AND FORMATION RATE}\label{sec:function}
In this section, we calculate the luminosity evolution and formation rate in X-ray for FXTs using a nonparametric statistical method, which operates under the assumption that the two variables are independent or uncorrelated — here $L$ and $z$, which means the assumption of no cosmological luminosity evolution.
Currently, a widely utilized approach for assessing the correlation between luminosity and redshift is the Efron-Petrosian method, developed by \citet{1992ApJ...399..345E}. If a correlation is identified, a new variable $g(z)$ (the evolution function) is introduced to ensure that the relationship between luminosity ( $L_0 = L/g(z)$ ) and redshift $z$ becomes uncorrelated. This allows for the application of nonparametric methods, similar to the Lynden-Bell's $c^-$ method \citep{10.1093/mnras/155.1.95}, to determine the intrinsic luminosity and redshift distributions. Those methods have been widely used in different objects, such as quasars\citep{10.1093/mnras/155.1.95,1992ApJ...399..345E,desai2016pulsar,2021ApJ...913..120Z,2024Univ...10..340Y}, LGRBs \citep{Lloyd-Ronning_2002,2012MNRAS.423.2627W,Petrosian_2015,2015ApJS..218...13Y,10.1093/mnras/stz2155}, short GRBs \citep{Yonetoku_2014,2018ApJ...852....1Z,2020ApJ...896...83G}, galaxies \citep{1978AJ.....83.1549K,1986ApJ...307L...1L,10.1093/mnras/221.2.233}, and fast radio bursts (FRBs: \cite{DENG20191,2024ApJ...973L..54C,2025A&A...698A..18Z}.)

\subsection{luminosity evolution}
\begin{figure*}[ht!]
\plottwo{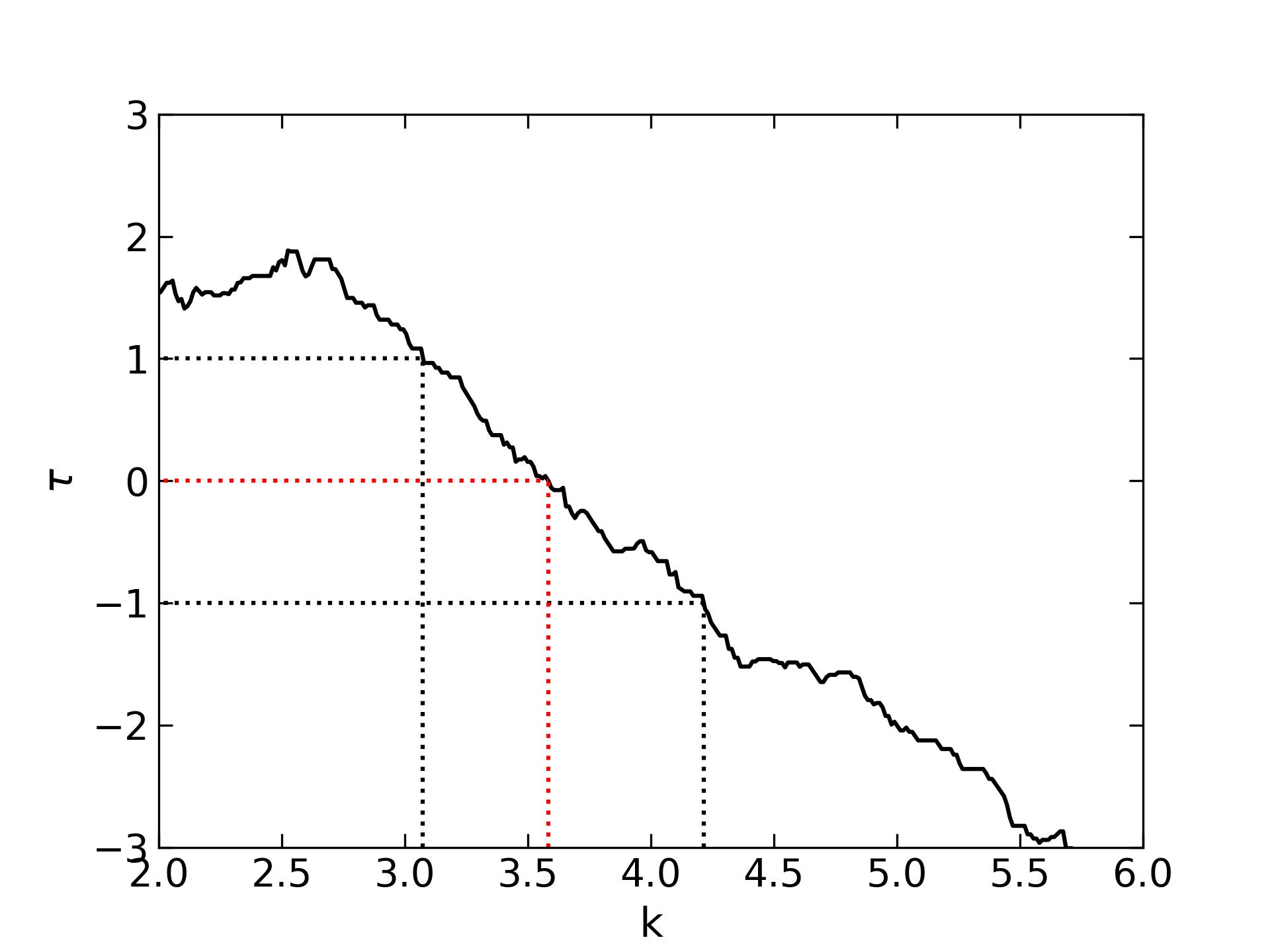}{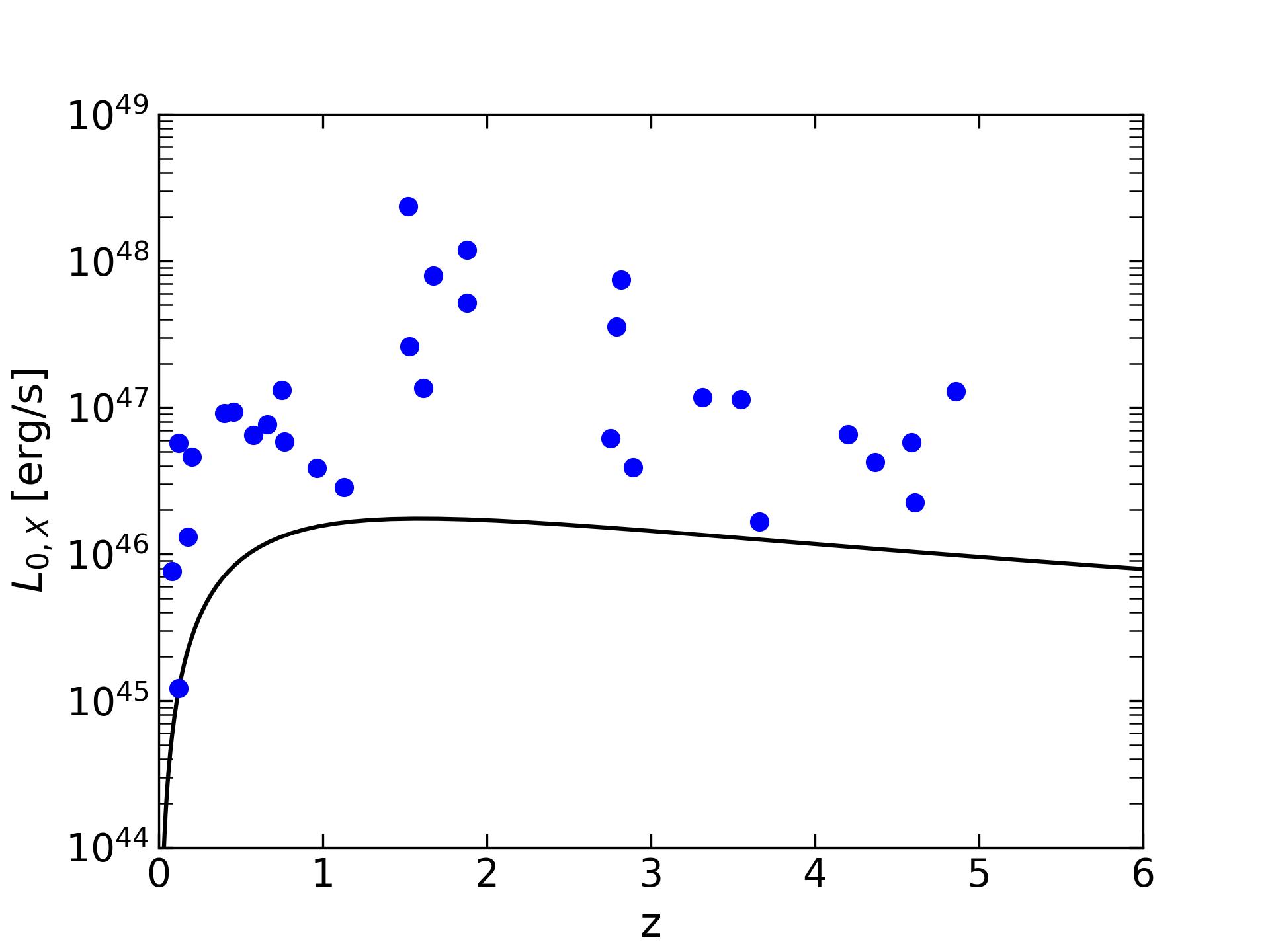}
\caption{Left: value of test statistic $\tau$ as a function of $k$. The red dotted line is the best fit for $\tau = 0$, and the black dotted lines represent 1$\sigma$ errors. The optimal  value is $k=3.58_{-0.51}^{+0.63}$ at a 1$\sigma$ confidence level. Right: nonevolving luminosity $L_{0,X} = L_{X}/(1 + z)^{3.58}$ of 31 FXTs.  The black solid line represents the observational limit considering the impact of $k$.
\label{tau}}
\end{figure*}

Figure \ref{lum_z} illustrates a strong correlation between luminosity and redshift, largely influenced by biases arising from data truncations. The Efron-Petrosian method addresses this issue using rank statistics based on a modified version of the Kendall $\tau$ test, defined as follows:
\begin{equation}
\tau = \frac{\sum_i(R_i-E_i)}{\sqrt{\sum_i V_i}}\,.
\end{equation}
Here, $R_i$ represents the normalized ranks of sources within their associated sets (comprising sources with $z_j < z_i$ and  $L_j >L_{\rm i,lim}$ for ranking in $L$). A detailed description of this test can be found in \cite{Lloyd-Ronning_2002,Yonetoku_2014,2015ApJS..218...13Y,Petrosian_2015}.
If luminosity and redshift are independent, the $R_i$ values should be uniformly distributed between 1 and $n_i$. Consequently, the expected mean and the variance of $R_i$ are given by $E_i=\frac{n_i+1}{2}$ and $V_i=\frac{n_i^2-1}{12}$, respectively. 
To quantify the correlation and the luminosity evolution function, we utilize a simple power-law model given by $g(z) = (1 + z)^{k}$. Theoretically, $|\tau(k=0)| >1$ indicates a strong dependence between luminosity and redshift, while $|\tau| = 0$ suggests independence. The left panel of Figure \ref{tau} presents $\tau$ as a function of $k$, revealing that the best fit occurs at $k=3.58_{-0.51}^{+0.63}$ with a 1$\sigma$ confidence level.
Additionally, we display the distribution of nonevolving luminosity and redshift in the right panel of Figure \ref{tau}.

\subsection{Luminosity Function}
After removing the evolution's effect through $L_0 = L/(1+z)^{3.58}$, we can derive the local cumulative luminosity function $\psi(L_0)$ with the Lynden-Bell's $c^-$ method from the following equation \citep{10.1093/mnras/155.1.95}:
\begin{equation}
\Psi(L_{0i}) = \prod_{j<i}(1+\frac{1}{N_j})\;,
\end{equation}
where $N_j$ represents the number of objects in the new associated set ($N_i$, blue rectangle) of FXT $i$, and $j<i$ means the $j_{th}$ FXT has a larger luminosity than the $i_{th}$ FXT. We employ a broken power-law function to model the cumulative luminosity function. The best-fit parameters are provided by
\begin{equation}
\Psi(L_{0})\propto 
\begin{cases}
L_0^{-0.25\pm 0.01}, & L_0 \leq L_0^b \\
L_0^{-0.71\pm 0.03}, & L_0>L_0^b \\
\end{cases}
\end{equation}
where the break luminosity $L_0^b$=$(4.17 \pm 0.34) \times 10^{46}$ erg/s. The cumulative luminosity distribution and the fitted function are displayed in the left panel of Figure \ref{cum}. During the fitting process, we employ a fitting method that accounts for errors, assuming that the errors for each data point follow a Poisson distribution. It is necessary to note that this is the luminosity function at $z=0$, because the luminosity evolution is removed. The differential luminosity function can be directly obtained by $\psi(L,z)=\psi(L/g(z))=-\frac{d\Psi(L_0)}{dL_0}$ with $L=L_0(1+z)^k$. The derivatives of this function provide a comparable characterization for the differential luminosity function. For instance, the indices $-0.25$ and $-0.71$ of $\Psi(L_{0i})$ indicate that $\psi(L/g(z))$ can also be fitted to a broken power law with indices $-1.25$ and $-1.71$. 

\subsection{Formation Rate Evolution}
\begin{figure*}[ht!]
\plottwo{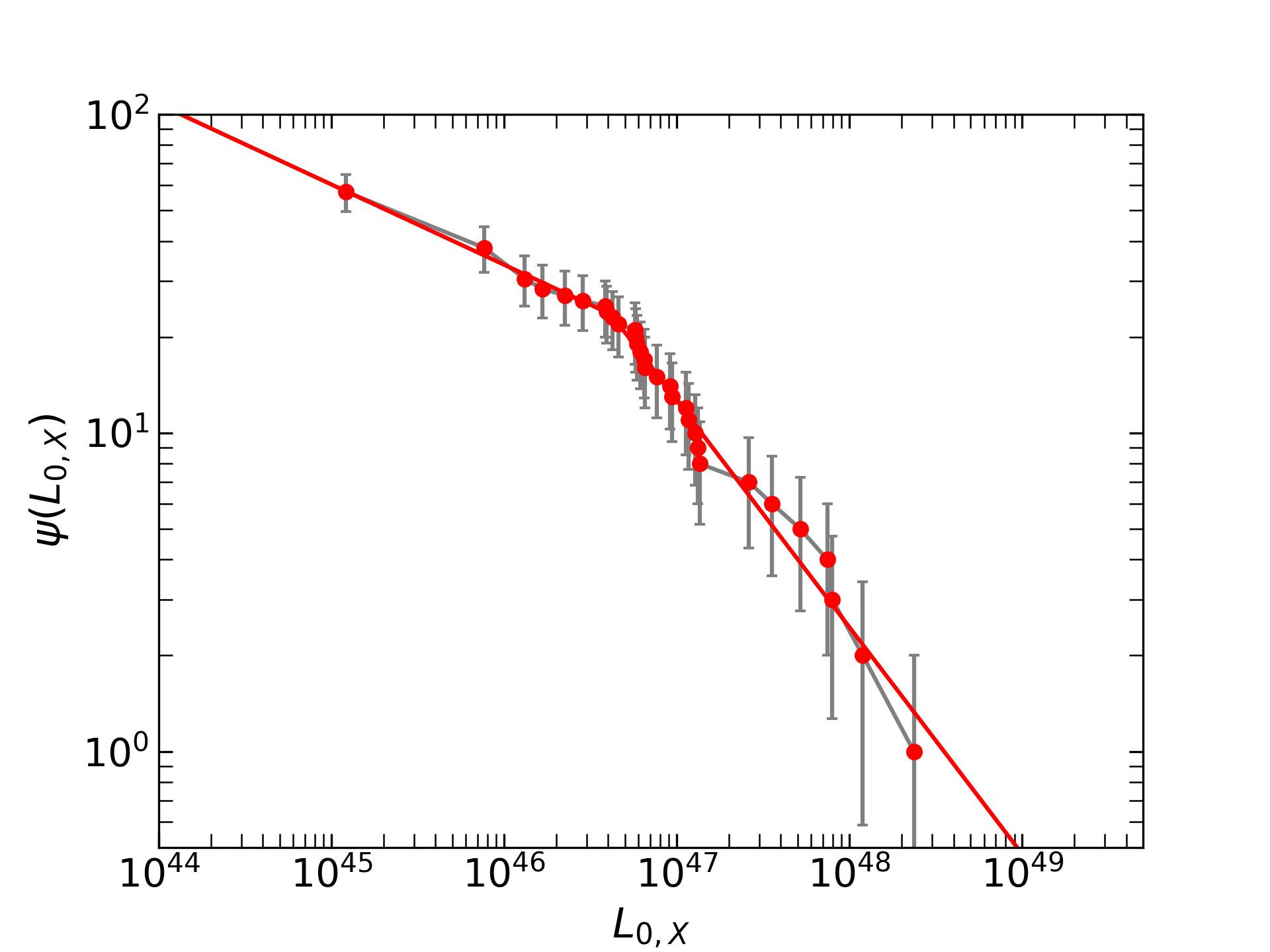}{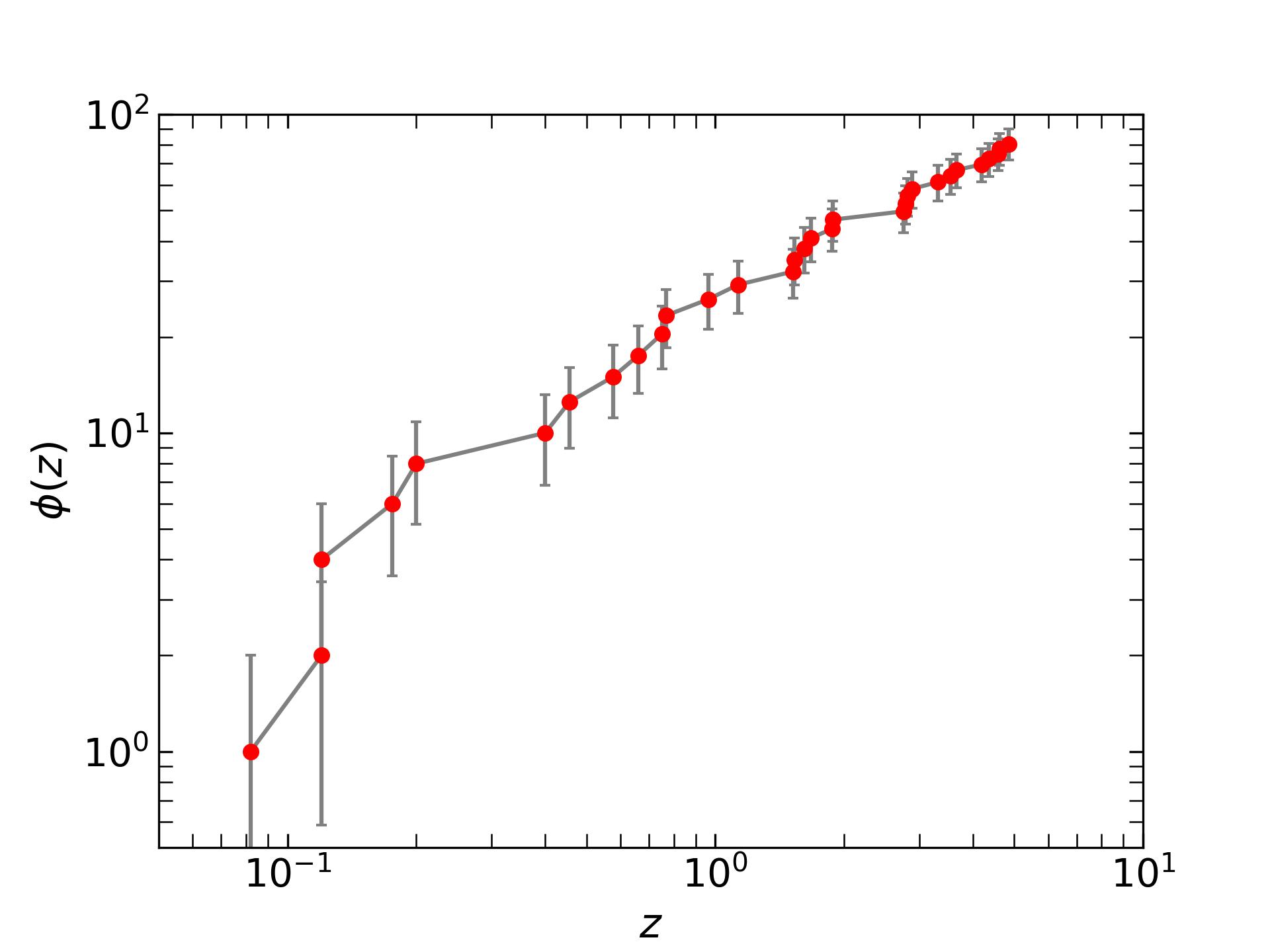}
\caption{Left: cumulative luminosity function $\psi(L_{0,X})$. The scatter represents the cumulative number (N) with 1$\sigma$ error ($\sqrt{N}$), and the red line represents the best fit with a broken power-law model with the low-end index of $0.25 \pm 0.01$ and the high-end index of $0.71 \pm 0.03$, and the break luminosity is $4.17 \pm 0.34 \times 10^{46}$ erg/s. Right: cumulative redshift distribution of FXTs, $\phi(z)$.
\label{cum}}
\end{figure*}

The cumulative redshift distribution $\phi(z)$ can be obtained from
\begin{equation}
\phi(z_i) = \prod_{j<i}(1+\frac{1}{M_j})\;,
\end{equation}
where $M_j$ represents the number of objects in the new associated set ($M_i$, red rectangle) of FXT $i$, and $j<i$ means the $j_{th}$ FXT has a smaller redshift than the $i_{th}$ FXT. The right panel of Figure \ref{cum} shows the cumulative redshift distribution $\phi(z_i)$ of FXTs. The formation rate of FXTs can be derived from
\begin{equation}
\rho(z)=
\frac{d\phi(z)}{dz}(1+z)
\left(\frac{dV(z)}{dz}\right)^{\!-1},
\label{eq:rho}
\end{equation}
where $(1+z)$ results from the cosmological time dilation and $dV(z)/dz$ is the differential comoving volume, which can be expressed as:
\begin{equation}
\frac{dV(z)}{dz}
=4\pi\left(\frac{c}{H_0}\right) 
\frac{d_L(z)^2}{(1+z)^2 \sqrt{1-\Omega_m+\Omega_m(1+z)^3}},
\label{eq:dV}
\end{equation}

\begin{figure*}[ht!]
\plottwo{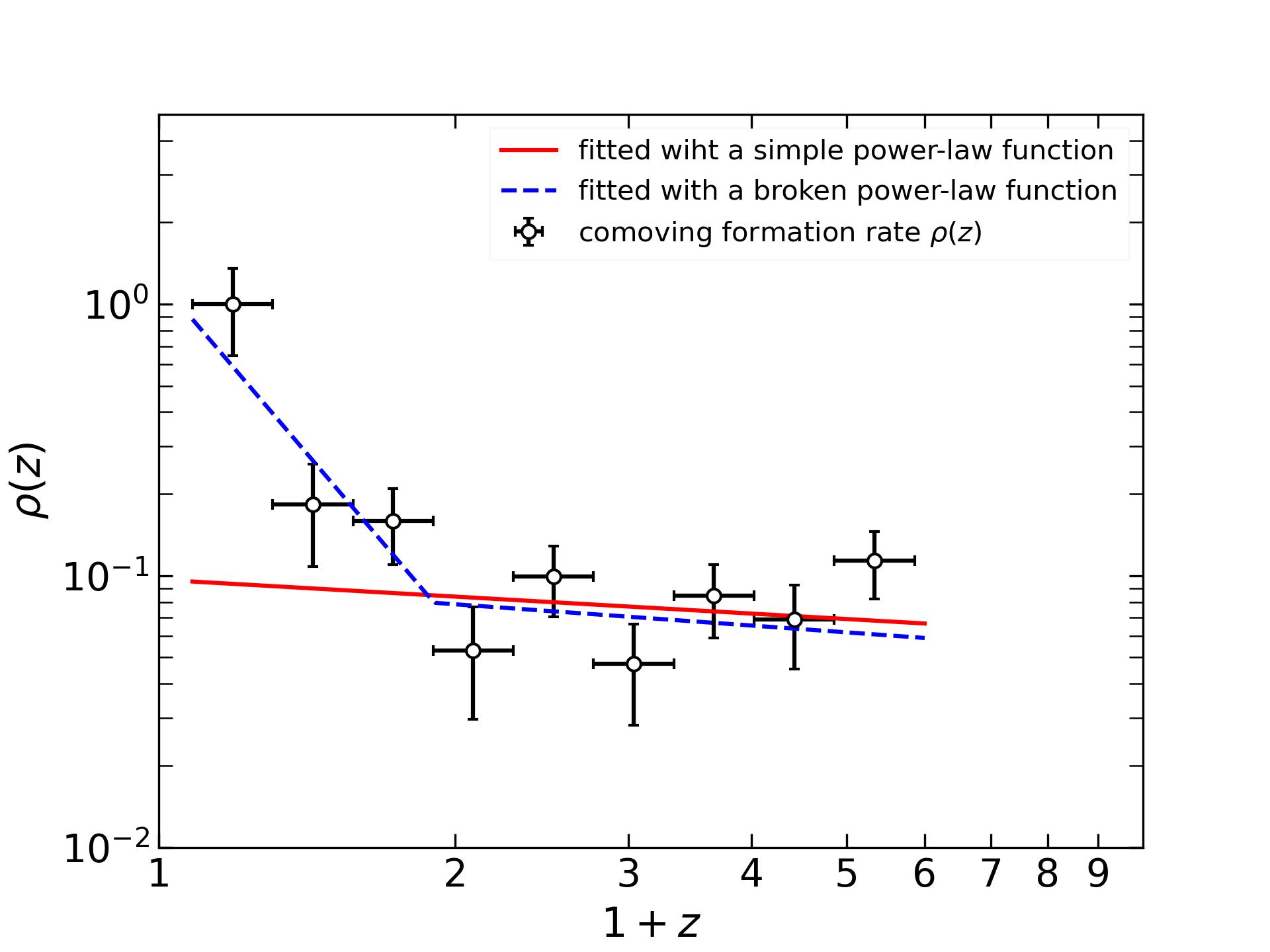}{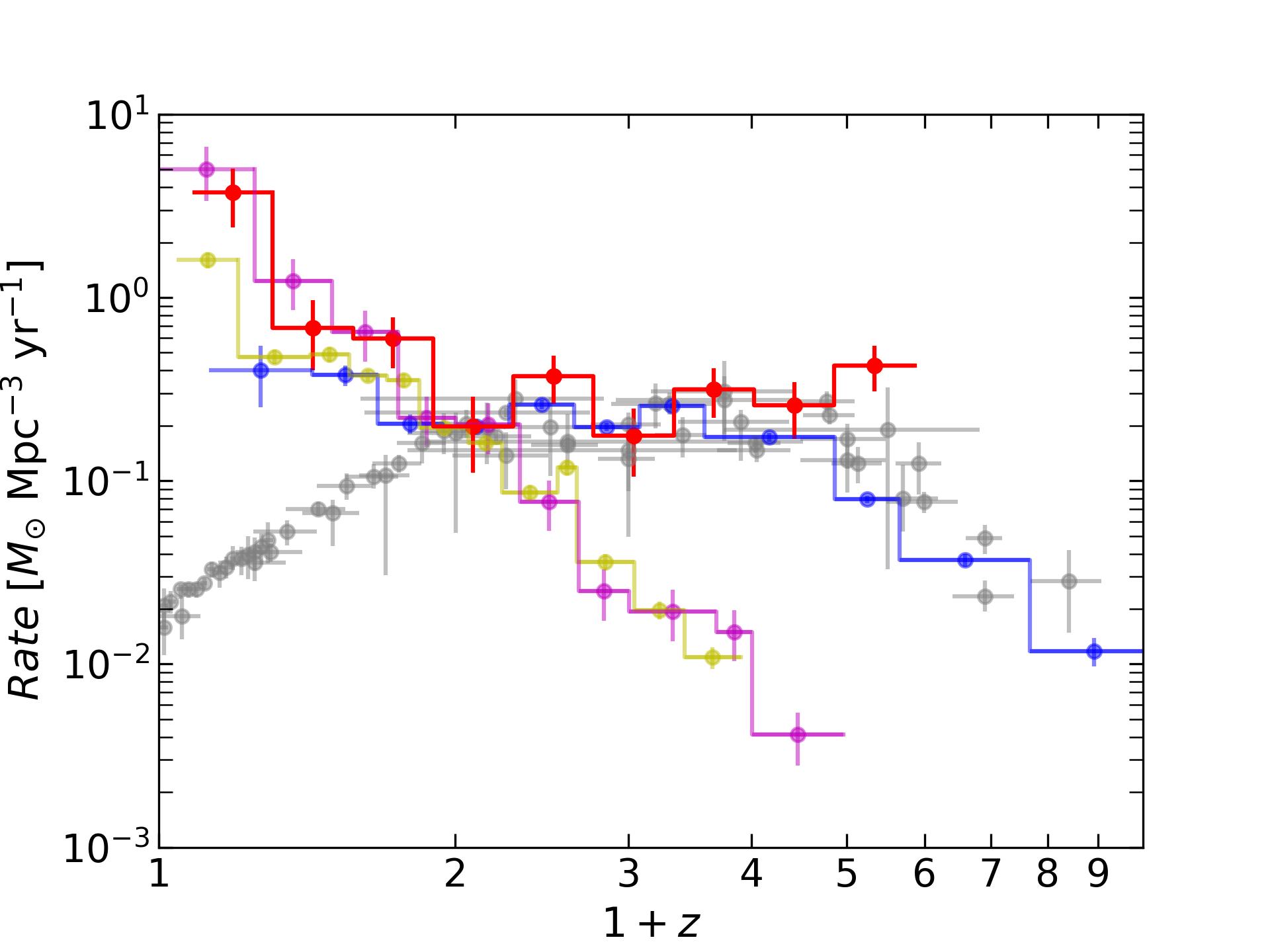}
\caption{Left: comoving formation rate $\rho(z)$ of FXTs obtained from Equation (\ref{eq:rho}). It is normalized to unity at the first point, and the 1$\sigma$ error is also shown. The solid and dashed lines represent the single and double power-law forms of $\rho(z)$, respectively. Right: comparison between the formation rate of FXTs and other events. The red line represents the formation rate of FXTs. The yellow and blue lines represent the rates of short GRBs \citep{2018ApJ...852....1Z} and LGRBs \citep{2015ApJS..218...13Y}, respectively. The magenta line is the rates of FRBs \citep{2024ApJ...973L..54C}. The gray dots correspond to the observed SFR \citep{Hopkins_2006,2008MNRAS.388.1487L}. Note that all formation rates are normalized at a redshift of $\sim 1$
\label{rate}}
\end{figure*}

The left panel of Figure \ref{rate} gives the normalized comoving formation rate $\rho(z)$ of FXTs, and the error bar gives a $1\sigma$ confidence level. The differential distribution $\frac{d\phi(z)}{dz}$ is numerically derived by applying a finite-difference method to the cumulative redshift distribution shown in the right panel of Figure 3.
We model the FXT formation rate using two distinct functional forms: a simply power-law function $\rho(z)= \rho(0)(1+z)^{-\delta}$ and a broken power-law function:
\begin{equation}
\rho(z) = \rho(0)
\begin{cases}
(1+z)^{-\gamma_1}, & z \leq z_b \\
(1+z_b)^{\gamma_2-\gamma_1}(1+z)^{-\gamma_2}, & z>z_b \\
\end{cases}.
\end{equation}
Given the physical plausibility and the substantial uncertainties in the observational data at $z>1.0$, we fix the parameter $\gamma_2 = -0.26$ to the value of the star formation rate (SFR) slope for $0.97<z<4.48$ reported by \cite{Hopkins_2006}. The remaining parameters are fitted to the data in the left panel of Figure \ref{rate}, yielding $\delta = 0.21 \pm 0.59$ for the single power-law model and $\gamma_1 = 4.25 \pm 2.38$ with $z_b = 0.90 \pm 0.33$ for the broken power law model. A comparison of the best-fit curves (solid and dashed lines) shows that the broken power-law provides a better description of $\rho(z)$.
Moreover, the local formation rate $\rho(0)$ can be estimated by the expected number of FXTs detected by EP, which can be expressed as
\begin{equation}
N=\frac{\Omega T}{4\pi}\int_{z_{\rm min}}^{z_{\rm max}}dz\int_{{\rm max}\left[L_{\rm min},\,L(F_{\rm limit},z)\right]}^{L_{\rm max}}\frac{\psi(L,z)\rho(z)}{1+z}\frac{dV}{dz}dL,
\label{eq:Number}
\end{equation}
where $\Omega = 1.1$ sr corresponds to the field of view (FoV) of 3600 ${\rm deg^2}$ of WXT, $T \sim 1.64$ yr is the observation period of $\sim 600$ days (from 2024 January 9 to 2025 August 31) that covers the sample of 107 FXTs. Here,  $z_{\rm min}=0$ and $z_{\rm max}=10.0$ are adopted, and the luminosity function is assumed to extend between $L_{\rm min}= 10^{45}$ erg/s and $L_{\rm max}=10^{52}$ erg/s. Note that the normalization factor of the luminosity function $\psi(L,z)$ is determined by requiring $\int_{L_{\rm min}}^{L_{\rm max}}\psi(L,z=0)dL=1$. By substituting the values obtained above into Equation (\ref{eq:Number}), we obtain a local formation rate of $\rho(0)=28.0_{-13.7}^{+26.6}$ Gpc$^{-3}$ yr$^{-1}$ and $153.8_{-95.1}^{+249.4}$ Gpc$^{-3}$ yr$^{-1}$, which correspond to a power-law distribution and a broken power-law distribution, respectively. Note that the uncertainty of this value is derived from the one-sigma confidence interval of the fitting parameters for $\rho(z)$.

\section{CONCLUSION AND DISCUSSION} \label{sec:conclusion}
In this study, we have employed the rigorous nonparametric methods of the Efron-Petrosian and Lynden-Bell methods to investigate the luminosity function and formation rate of FXTs based on the latest catalog of EP sources provided by \citet{Wu2025_inprep}. First, we determine the luminosity evolution using the Efron-Petrosian procedure proposed by \citet{1992ApJ...399..345E}, which enables us to define a deevolved (local) luminosity ($L_0$) that is independent of redshift ($z$). This approach facilitates the application of the Lynden-Bell's $c^-$ method for assessing the univariate distributions of $L_0$ and $z$.
Our results can be summarized as follows:
\begin{itemize}
    \item 
    Our analysis indicates that FXTs exhibit significant luminosity evolution up to redshift $z<5.0$, which can be approximated by the relationship $L \propto (1+z)^{3.58}$. This result is analogous to the luminosity evolution observed in other source populations; for example, the evolution factor for LGRBs is approximately 2.5 \citep[e.g][]{2016A&A...587A..40P,2015ApJS..218...13Y,Petrosian_2015}, while for short GRBs, it is around 4.5 \citep[e.g][]{2018ApJ...852....1Z,2018MNRAS.477.4275P,2020ApJ...896...83G}. Similarly, FRBs show an associated value of 5 \citep{2024ApJ...973L..54C}. 
    
    \item 
    The cumulative luminosity function, denoted as $\Psi(L_0)$, exhibits a monotonic decrease and is well described by a broken power law. The slopes of this function are characterized by $-0.25 \pm 0.01$ for low luminosities and $-0.71 \pm 0.03$ for high luminosities, with a characteristic break luminosity at $L_0^b =(4.17 \pm 0.34) \times 10^{46}$ erg/s. The slopes are consistent with those obtained by \citet{2012MNRAS.423.2627W,2015ApJS..218...13Y,2016A&A...587A..40P} for LGRBs. Notably, the break luminosity identified here is lower than that associated with LGRBs, suggesting that FXTs may represent a subset of LGRBs, specifically low-luminosity LGRBs.
 
    \item 
    The formation rate of FXTs has been obtained, with results presented in the left panel of Figure \ref{rate}. When described by a single power law, the index is found to be $-0.21 \pm 0.59$. However, it appears that within the redshift range of $1 < z < 5$, the rate remains approximately constant. In contrast, for $z < 1$, there is a rapid decline that deviates from the typical SFR\citep{Hopkins_2006}. This finding is consistent with the formation rate of LGRBs, which also shows an excess at low redshift \citep[e.g][]{2015ApJS..218...13Y,Petrosian_2015,10.1093/mnras/stz2155}.
    In the right panel of Figure \ref{rate}, we present a comparative analysis of the formation rate of FXTs in relation to the observed SFR and the occurrence of GRBs. It is important to note that all formation rates are normalized at a redshift of 1. 
    The FXT formation rate tracks the SFR and the LGRB rate at $z>1$. At $z<1$, however, it deviates from these trends and instead aligns with the rates of SGRBs and FRBs. This distinct evolutionary pattern indicates that the FXT population cannot be explained by a single progenitor model.
    
    \item 
    We have determined the local FXT rate as $\rho(0)=153.8_{-95.1}^{+249.4}$ Gpc$^{-3}$ yr$^{-1}$. This result is base aligned with the estimated formation rates of low-luminosity LGRBs, which range from approximately 100 to 700 Gpc$^{-3}$ yr$^{-1}$ \citep{2015ApJ...812...33S,2006Natur.442.1011P,2007ApJ...662.1111L}. Our derived rate is slightly higher than the value of $\rho(0)=29.2_{-24}^{+67.2}$ Gpc$^{-3}$ yr$^{-1}$ reported by \citet{2025arXiv250417034L} from observations of the extremely soft and weak FXT EP250108a. Furthermore, our derived rate is higher than the rates 
  for LGRBs, which are estimated to be about $1-30$ Gpc$^{-3}$ yr$^{-1}$ \citep{2010MNRAS.406.1944W,2015ApJS..218...13Y,2019MNRAS.488.4607L,2024MNRAS.532.3926K}. However, our results are significantly lower than the formation rate for FRBs, which is on the order of $\propto 10^4$ Gpc$^{-3}$ yr$^{-1}$ \citep{2024ApJ...973L..54C,2025A&A...698A..18Z}.

\end{itemize}

The physical origins of FXTs remain poorly understood; our research suggests a potential correlation with LGRBs. 
This conclusion is consistent with the comparative analysis conducted by \citet{2025arXiv250907141O}, which examined the cumulative redshift distributions of FXTs and long GRBs, revealing no statistically significant differences between these two populations.
It should be noted that the Efron-Petrosian method achieves higher statistical power with larger sample sizes. Although our sample of 31 FXTs represents the largest such catalog to date, the conclusions drawn should be treated with caution due to this limitation.

\begin{acknowledgments}
This work is based on the data obtained with Einstein Probe, a space mission supported by the Strategic Priority Program on Space Science of Chinese Academy of Sciences, in collaboration with the European Space Agency, the Max-Planck-Institute for Extraterrestrial Physics (Germany), and the Centre National d'Études Spatiales (France). This work is supported by the Strategic Priority Research Program of the Chinese Academy of Sciences (grant No. XDB0550400), the National Key R\&D Program of China (2024YFA1611704),
and the National Natural Science Foundation of China (grant Nos. 12321003, 12473049, 12422307, 12373053 and 12225305).
\end{acknowledgments}



\bibliography{ms}{}
\bibliographystyle{aasjournalv7}

\end{document}